\documentclass{article}

\usepackage{mathptmx}
\usepackage{arxiv}
\usepackage{graphicx}
\usepackage{hyperref}      
\usepackage{amsmath,amsfonts,amssymb}
\usepackage{color}
\usepackage{braket}
\usepackage{float}
\usepackage{subfig}
\usepackage{cite}

\hypersetup{
    colorlinks=true,
    linkcolor=red,
    filecolor=cyan,      
    urlcolor=magenta,
    citecolor=blue}

\title{Propagation of light in linear and quadratic GRIN media: The Bohm potential}
\author{Felipe A. Asenjo
\\ Facultad de Ingeniería y Ciencias, Universidad Adolfo Ibáñez, Santiago 7491169, Chile.
  \\ \And Sergio A. Hojman\\
  Departamento de Ciencias, Facultad de Artes Liberales,
Universidad Adolfo Ibáñez, Santiago 7491169, Chile.\\
Departamento de Física, Facultad de Ciencias, Universidad de Chile,
Santiago 7800003, Chile.\\
Centro de Recursos Educativos Avanzados,
CREA, Santiago 7500018, Chile.
\\  \And Héctor M. Moya-Cessa and Francisco Soto-Eguibar \\
 Instituto Nacional de Astrofísica Óptica y Electrónica\\Calle Luis Enrique Erro No. 1, Santa María Tonantzintla, Pue., 72840, Mexico.\\}

\begin{document}

\maketitle

\begin{abstract}
It is shown that field propagation in linear and quadratic gradient-index (GRIN) media obeys the same rules of free propagation in the sense that a field propagating in free space has a (mathematical) form that may be {\it exported} to those particular GRIN media. The Bohm potential is introduced in order to explain the reason of such behavior: it changes the dynamics by modifying the original potential. The concrete cases of two different initials conditions for each potential are analyzed.
\end{abstract}

\section{Introduction}
Quadratic gradient-index (GRIN) media has been studied extensively over the years \cite{Bandres2007,Arrizon2015,Wiersma2016,Peng_2019,Dong2020,Zhu_2020}. It is well-known that such media produce focusing of the incoming field as it generates its fractional Fourier transform that for certain distances of propagation produces the full Fourier transform. Beyond paraxial propagation, it has been shown that it may be used as a beam splitter \cite{Arrizon2015}.\\
It is said  that an Airy beam propagating in free space accelerates as a function of the propagation distance \cite{Siviloglou2007,Vo2010,Efremidis2019}. This effect can be fully canceled by propagating such an Airy beam in a linear potential, or even the direction of the light bending may be adjusted arbitrarily \cite{ChavezCerda2011}. It is, therefore, a well-known fact that Airy beams keep their structure (they are Airy beams during propagation) in both free propagation and linear potential. As Airy beams are unphysical in the sense that they would carry infinite energy, some modifications have been proposed that produce interesting properties as they propagate, among them self-focusing \cite{Jauregui2014,Torre2015,Qian2017,Wang2019,Marsal2019}. Effects of GRIN media on Airy beam propagation have been also studied \cite{Peng_2019,Dong2020}.\\
In this contribution we show that Airy beams also keep their structure during propagation in a quadratic potential  (except for propagation distances $z=\frac{(2m+1)\pi}{2}$ where the fractional Fourier transform becomes the complete Fourier transform). Airy beams are unphysical in the sense that they have infinite energy; in order to produce beams with finite energy (so that the wave function be normalizable) and physically feasible, we multiply the Airy function by an exponential factor ($e^{\gamma x}$ or $e^{-\gamma x^2}$). We show that even with these factors the modified Airy beams keep their structures, although they acquire a complex argument. We argue that the Bohm potential may be used as an explanation of the beam bending as it modifies the initial potential \cite{Asenjo12017,Hojman12020,Hojman22020}.\\ 

\section{Propagation of structured beams in linear and quadratic GRIN media: Linear and quadratic}
It may be easily shown that field propagation in linear and quadratic GRIN media obeys the same rules of free propagation, in the sense that a field propagating in free space has a (mathematical) form that may be {\it exported} to those particular GRIN media. In units such that the wavevector $k$ is set as one, we write the paraxial propagation in free space as \cite{Siegman1986})
\begin{equation}\label{ecsch}
i \frac{\partial \psi}{\partial z}=-\frac{1}{2}\frac{\partial^2 \psi}{\partial x^2}.
\end{equation}
By defining $\hat{p}=-i\frac{\partial}{\partial x}$ and $\hat{x}=x$, the formal solution for this paraxial equation reads
\begin{equation}\label{forsolecsch}
\psi(x,z)=e^{-i\frac{z\hat{p}^2}{2}}\psi(x,0),
\end{equation}
which may be proved by  deriving the above equation to recover (\ref{ecsch}).
\subsection{Linear interaction}
In the case of a linear GRIN medium, the paraxial equation is
\begin{equation}\label{ecschlin}
i\frac{\partial \psi}{\partial z}=\left(\frac{\hat{p}^2}{2}+\eta \hat{x}\right)\psi,
\end{equation}
and its propagator
\begin{equation}\label{propecschlin}
\hat{U}(z)=\exp \left[ -i z\left(\frac{\hat{p}^2}{2}+\eta \hat{x}\right)\right]
\end{equation}
can be written \cite{SotoEguibar2015} as
\begin{equation}\label{propecschlin2}
\hat{U}(z)=\exp\left(-i\frac{\eta^2z^3}{6}\right)
\exp\left(-i\eta z \hat{x} \right)
\exp\left(i\frac{\eta z^2}{2}\hat{p}\right)
\exp\left(-i z \frac{\hat{p}^2}{2}\right).
\end{equation}
We may see from the above equation that fields that maintain  their structure during free propagation, namely Bessel, Laguerre-Gauss, Hermite Gauss beams (all of them with a proper generalization to the 3-D case), will also keep their structure for the linear  potential. This fact is seen easily as the exponential that is further to the right has the same structure as (2). The first two exponential functions are simple phases. The third exponential displaces the solution by a quantity $\frac{\eta z^2}{2}$. It is straightforward to see that any field we propagate in free space exports the same structure of propagation to the linear case experiencing just a shift.

\subsection{Quadratic interaction}
A GRIN medium with quadratic behavior is equivalent to a harmonic oscillator. Propagation of light in such a medium obeys the paraxial wave equation 
\begin{equation}\label{ecschcua}
i\frac{\partial \psi}{\partial z}=-\frac{1}{2}\frac{\partial^2 \psi}{\partial x^2}+\frac{\eta^2x^2}{2}\psi.
\end{equation}
In Appendix \ref{apa}, we show that  the above equation has a propagator of the form
\begin{equation}\label{propecschcua}
\hat{U}(z)=\exp\left[-i\eta^2 f\left(z\right)\hat{x}^2\right]
\exp\left[-i\frac{g\left(z\right)}{2} \left( \hat{x}\hat{p}+\hat{p}\hat{x}\right) \right] 
\exp\left[-i f\left(z\right)\hat{p}^2\right],
\end{equation}
where
\begin{equation}\label{fg}
f\left(z\right)=\frac{\tan\left(\eta z \right)}{2\eta} ,\qquad
g\left(z\right)=\ln\left[\cos\left(\eta z \right)  \right]
\end{equation}
and which is in fact the fractional Fourier transform.\\
Given that the operators $\hat{x}^2$, $\hat{x}\hat{p}+\hat{p}\hat{x}$ and $\hat{p}^2$ obey a simple algebra, the propagator may be written in terms of $2\times 2$ matrices \cite{Fisher1984,Siegman1986}, that although it is another way to find it, it does not clearly show that the propagation of an arbitrary field in free space will have the same behavior in a quadratic GRIN medium, as Eq. \eqref{propecschcua} does.\\
It is worth to note that the third exponential in the above propagator corresponds to free propagation by making $f(z)\rightarrow z/2$. The second exponential corresponds to the well-known effect of  squeezing in quantum mechanics, which basically  scales the (wave) function to which is applied (and a normalization function that depends on the propagation distance) and the last term is a simple quadratic phase. Note also that as the functions $f(z)$ and $g(z)$ are periodic,so we will have the complete recovery of the initial field at certain distances of propagation $z$.\\
Moreover, the structure of the propagators (\ref{propecschlin2}) and (\ref{propecschcua}) allows us to state that the free propagation solutions for arbitrary fields may be directly imported to the linear and quadratic potentials, by just adding a displacement (linear potential) or a scaling (quadratic potential). Therefore, in what follows, we show that the Airy beams also propagate keeping their structure while they propagate in a GRIN medium.

\section{The linear potential}
\begin{figure}[H]
\centering
\subfloat[Initial condition $\psi(x,z=0)=\exp(\gamma x)\textrm{Ai}(\epsilon x)$ with parameters $\gamma=0.25, \; \epsilon=2.0$ and $\eta=1.0$]
{\includegraphics[width=0.4\textwidth]{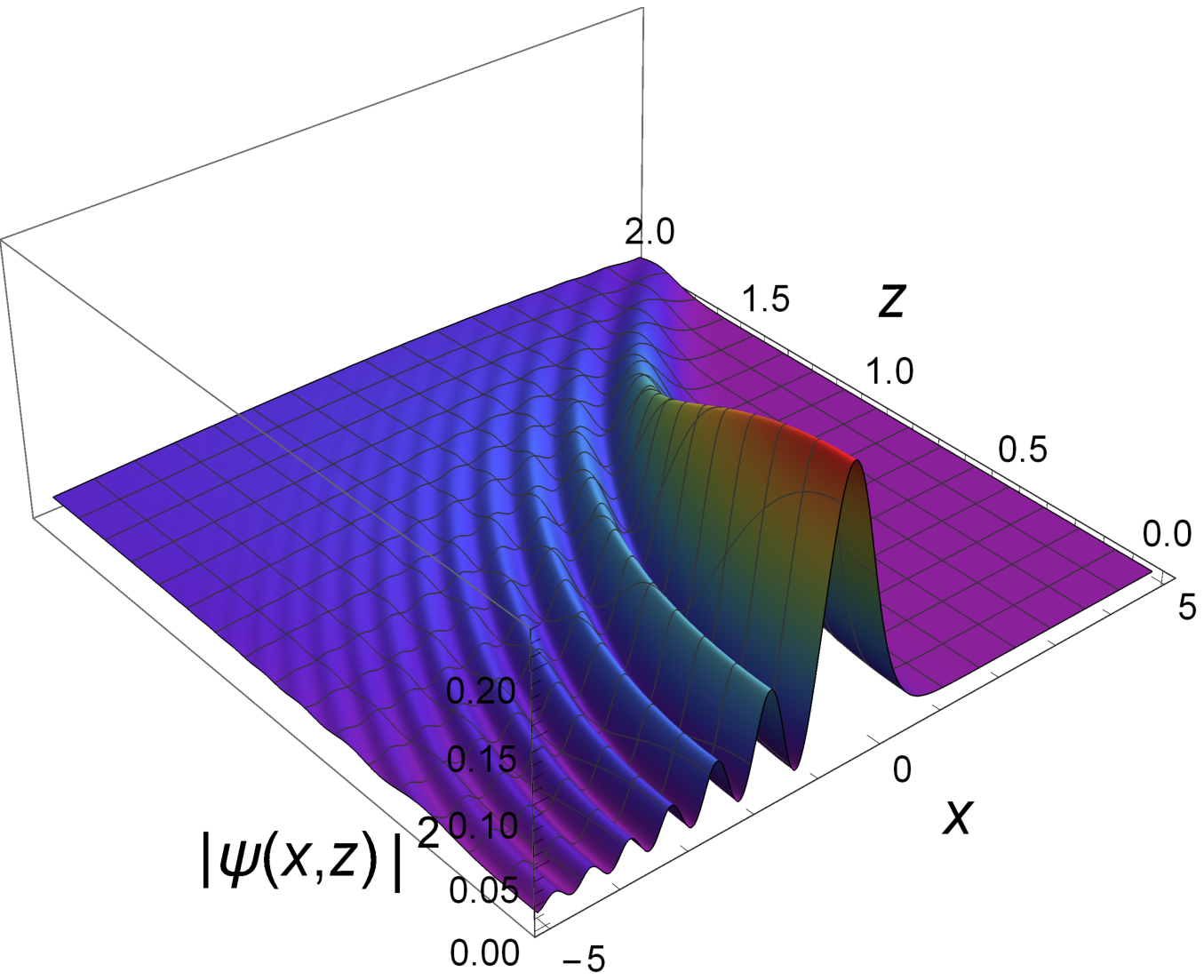}}
\hspace{14 pt}
\subfloat[Initial condition $\psi(x,z=0)=\exp(-\gamma x^2)\textrm{Ai}(\epsilon x)$ with parameters $\gamma=0.1, \; \epsilon=2.0$ and $\eta=2.0$]
{\includegraphics[width=0.4\textwidth]{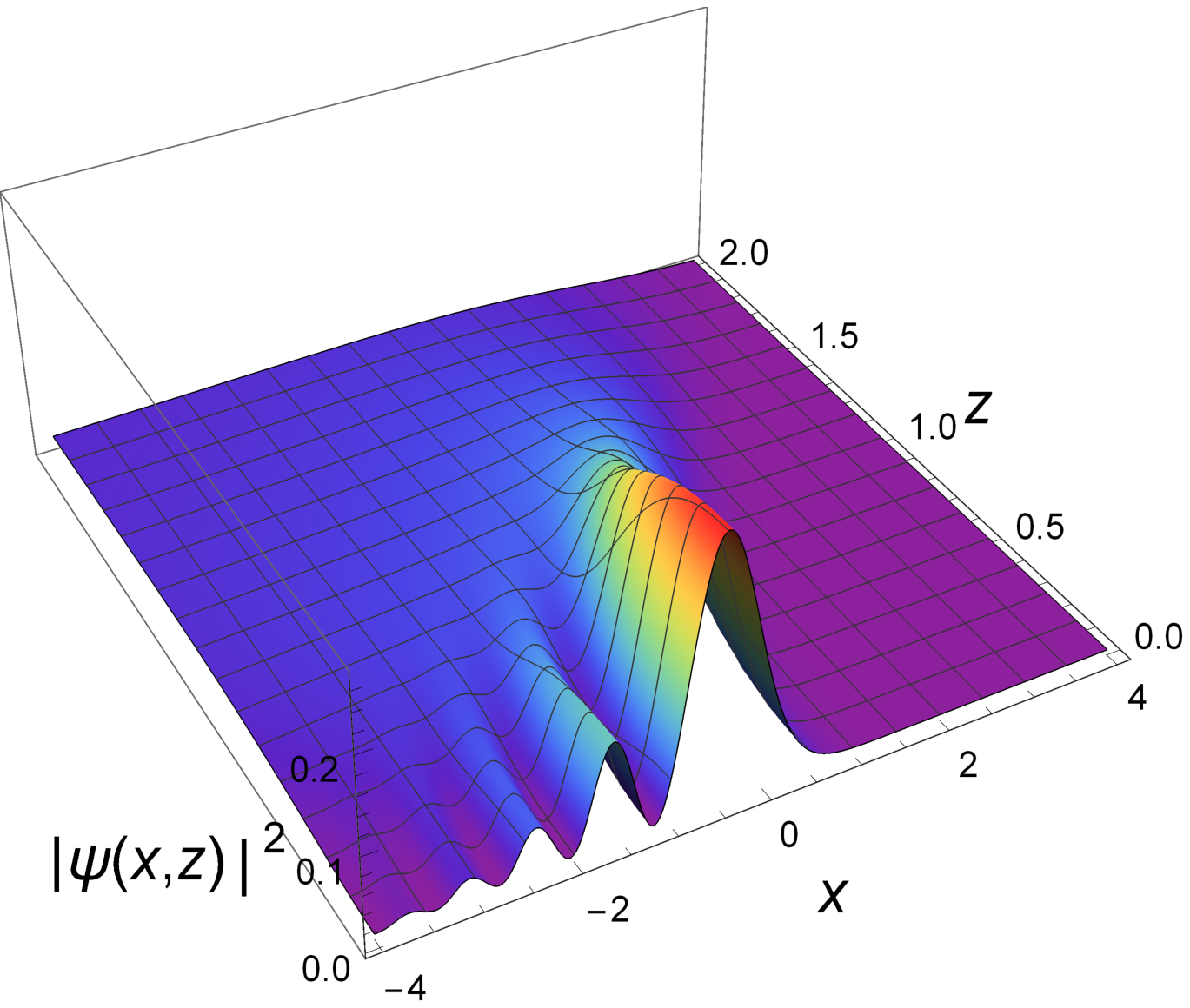}}
\caption{The intensity $|\psi(x,z)|^2$ for the linear potential with the two different initial conditions \label{3DIntLin}}
\end{figure}

In the case of the linear potential, we consider two different initial conditions. First, an exponential by an Airy function, $\exp\left( \gamma x\right) \mathrm{Ai}(\epsilon x)$, and second, a Gaussian multiplying an Airy function, $\exp\left(-\gamma x^2\right) \mathrm{Ai}(\epsilon x)$, where we assume that the parameters $\gamma$ and $\epsilon$ are always real and strictly positive. In both cases, we have a field with finite energy, as the calculation of the integral $\int_{-\infty}^{\infty}\left| \psi(x,z=0)\right|^2 dx$ is finite for both cases.\\
Applying the factorized evolution operator \eqref{propecschlin2} to the initial conditions and using the usual techniques of quantum optics, we find for the Exponential-Airy function that
\begin{align}\label{psilinexpai}
\psi(x,z)=&\exp\left\lbrace 
i  \left[\frac{\gamma^2 z}{2}
+\left(\frac{\varepsilon^3}{2}-\eta\right)xz
+\left(-\frac{\varepsilon ^6}{12}+\frac{\varepsilon ^3 \eta }{4}-\frac{\eta ^2}{6}\right) z^3
\right] \right\rbrace 
\exp\left[\gamma x +\frac{\gamma z^2 }{2}
\left(\eta-\varepsilon^3 \right) \right]
\nonumber \\ &
\mathrm{Ai}(\epsilon x+\frac{\epsilon\eta z^2}{2}-\frac{\varepsilon^4 z^2}{4}+i\epsilon\gamma z).
\end{align}
and for the Gaussian-Airy function we get
\begin{equation}\label{psilingauai}
\psi(x,z)=\exp\left[i\sigma_1\left(x,z\right)\right]
\exp\left[\sigma_2\left(z\right)x^2+\sigma_3\left(z\right)x+\sigma_4\left(z\right)\right]
\mathrm{Ai}\left(\delta_1\left(z\right)x+\delta_2\left(z\right)\right),
\end{equation}
where
\begin{subequations}
\begin{align}
\sigma_1\left(z\right)=&
-\frac{1}{2}\arctan(2\gamma z)
+\frac{2\gamma^2 z}{4\gamma^2 z^2+1}x^2
+\frac{z \left[\varepsilon ^3-2 \eta -16 \gamma ^4 \eta  z^4-4 \gamma ^2 z^2 \left(\varepsilon ^3+3 \eta \right)\right]}{2 \left(4 \gamma ^2 z^2+1\right)^2} x
\nonumber \\ &
-\frac{z^3 \left[3 \varepsilon ^3 \eta  \left(16 \gamma ^4 z^4-1\right)+\varepsilon ^6 \left(1-12 \gamma ^2 z^2\right)
+2 \eta^2 \left(\gamma^2 z^2+1\right) \left(1 +4 \gamma ^2  z^2\right)^2\right]}{12 \left(4 \gamma ^2 z^2+1\right)^3},\\
\sigma_2\left(z\right)=&-\frac{\gamma}{4 \gamma ^2 z^2+1},\\
\sigma_3\left(z\right)=&
-\frac{\gamma z^2 \left(-2 \varepsilon ^3+\eta +4 \gamma ^2 \eta  z^2\right)}{\left(4 \gamma ^2 z^2+1\right)^2},\\
\sigma_4\left(z\right)=&
-\frac{1}{4} \log \left(4 \gamma ^2 z^2+1\right)
+\frac{\gamma  z^4 \left[\varepsilon ^6 \left(8 \gamma ^2 z^2-6\right)+12 \varepsilon ^3 \eta \left(1 +4 \gamma ^2  z^2\right)-3 \eta^2 \left(1 +4 \gamma ^2 z^2\right)^2\right]}{12 \left(4 \gamma ^2 z^2+1\right)^3},\\
\delta_1\left(z\right)=&
\frac{\varepsilon}{4 \gamma ^2 z^2+1}
-i \frac{2 \gamma  \varepsilon z}{4 \gamma ^2 z^2+1},\\
\delta_2\left(z\right)=&
\frac{\varepsilon  z^2 \left(-\varepsilon ^3+2 \eta +4 \gamma ^2 \varepsilon ^3 z^2+8 \gamma ^2 \eta  z^2\right)}{4 \left(4 \gamma ^2 z^2+1\right)^2}
+i \frac{\gamma  \varepsilon  z^3 \left(\varepsilon ^3-\eta-4 \gamma ^2 \eta  z^2\right)}{\left(4 \gamma ^2 z^2+1\right)^2}.
\end{align}
\end{subequations}
In both cases, we have split the exponential to show that the first factor is just a phase (remember that we have assumed that $\gamma, \varepsilon, \eta$ are positive real numbers). These propagated fields satisfy both the initial conditions and the original paraxial equation, Eq.\eqref{ecschlin}.\\
In Fig. \ref{3DIntLin}, we show the intensity of the fields expressed in \eqref{psilinexpai} and \eqref{psilingauai} as functions of $x$ and $z$. 

\section{The quadratic potential}
\begin{figure}[H]
\centering
\subfloat[Initial condition $\psi(x,z=0)=\exp(\gamma x)\textrm{Ai}(\epsilon x)$ with parameters $\gamma=1.0, \; \epsilon=2.0$ and $\eta=1.0$]
{\includegraphics[width=0.45\textwidth]{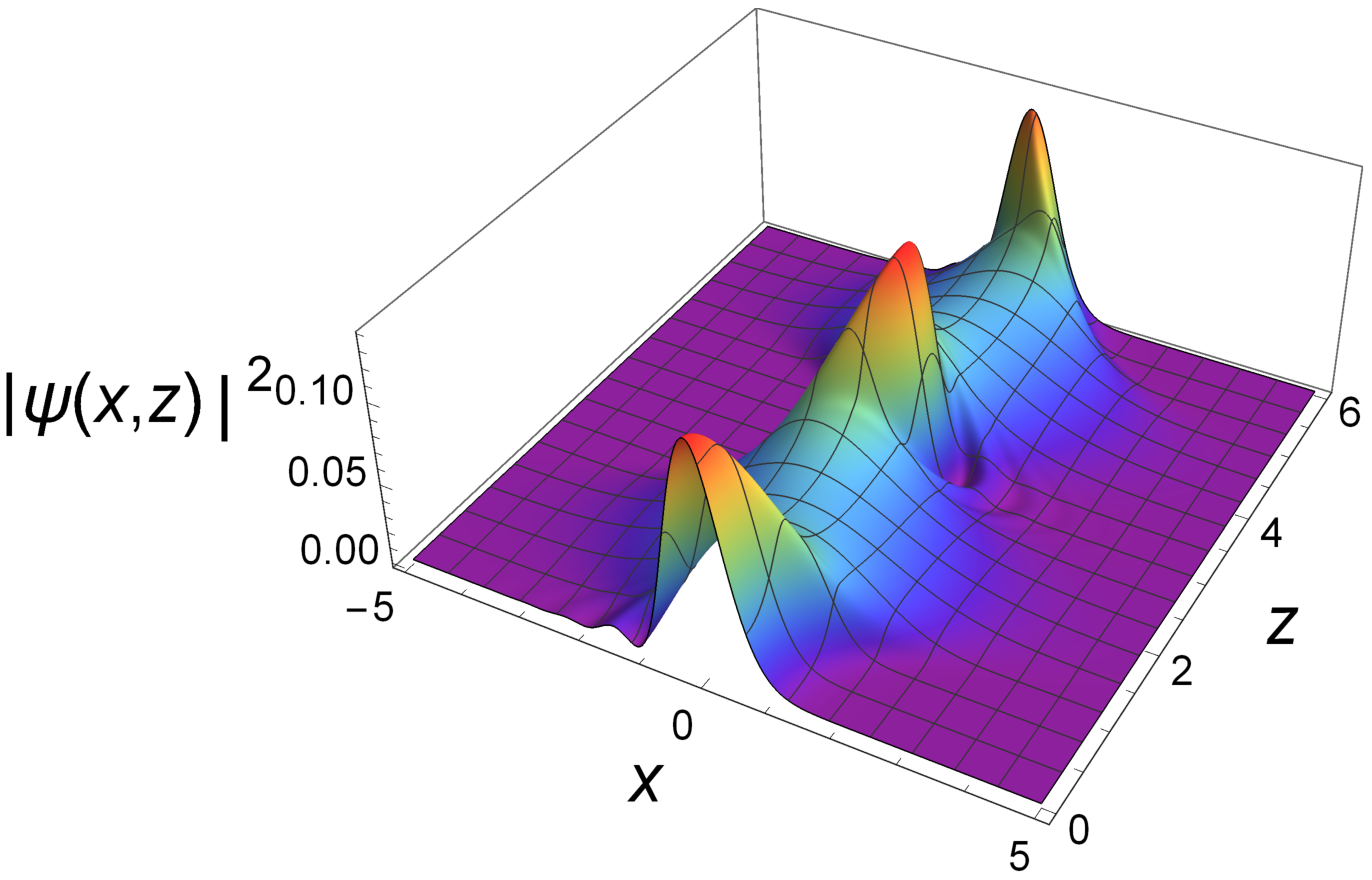}}
\hspace{14 pt}
\subfloat[Initial condition $\psi(x,z=0)=\exp(-\gamma x^2)\textrm{Ai}(\epsilon x)$ with parameters $\gamma=1.0, \; \epsilon=2.0$ and $\eta=1.0$]
{\includegraphics[width=0.45\textwidth]{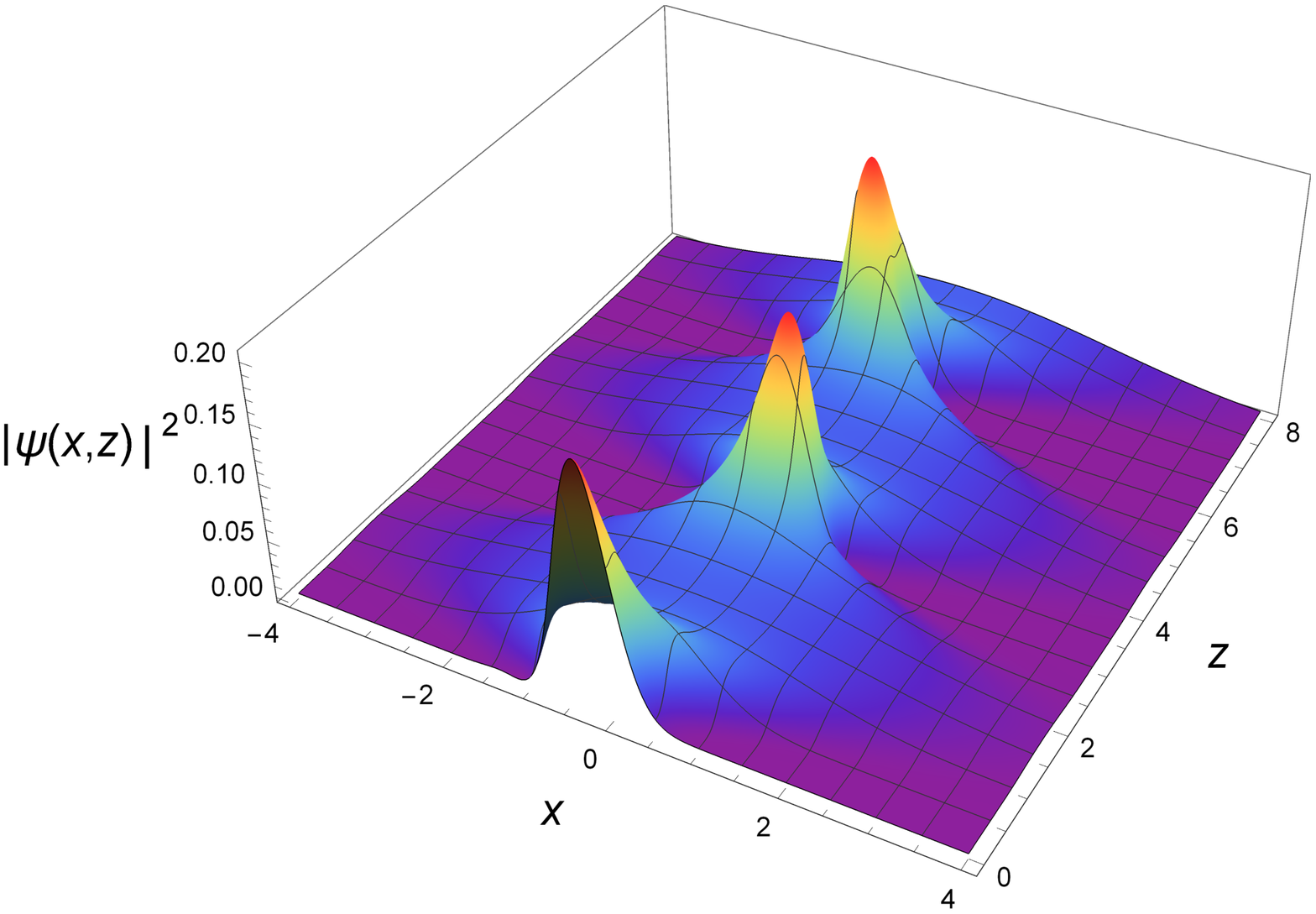}}
\caption{The intensity $|\psi(x,z)|^2$ for the quadratic potential with the two different initial conditions \label{3DIntCua}}
\end{figure}

In the case of the quadratic potential, we do the same thing that in the case of the linear potential. We use the same two initial conditions, an Exponential-Airy function and a Gaussian-Airy function, and we apply the evolution operator \eqref{propecschcua} to both initial conditions. For the Exponential-Airy function the propagated field, we find
\begin{align}\label{psicuaexpai}
\psi(x,z)=&
\exp i \left\lbrace i \frac{6 \eta ^2 \tan (\eta  z) \left[\gamma ^2-\eta ^2 x^2+\varepsilon ^3 x \sec (\eta  z)\right]-\varepsilon ^6 \tan ^3(\eta  z)}{12 \eta ^3}\right\rbrace
\nonumber \\ &
\exp \left\lbrace \gamma \sec (\eta  z) x -\frac{\gamma  \varepsilon ^3 \tan ^2(\eta  z)+\eta ^2 \log \left[ \cos (\eta  z)\right] }{2 \eta ^2}\right\rbrace
\nonumber \\ &
\mathrm{Ai}\left( \varepsilon \sec (\eta  z) x
-\frac{\varepsilon ^4 \tan ^2(\eta  z)}{4 \eta ^2}
+i\frac{\gamma \varepsilon  \tan (\eta  z)}{\eta }\right)
\end{align}
and for the the Gaussian-Airy function, we get
\begin{equation}\label{psicuagauai}
\psi\left(x,z \right)=\exp\left[i \tau_1\left(x,z\right) \right]  
\exp\left[\tau_2\left(z\right)x^2+\tau_3\left(z\right)x+\tau_4\left(z\right)\right]
\mathrm{Ai}\left(\rho_1\left(z\right)x+\rho_2\left(z\right)\right) 
\end{equation}
where
\begin{subequations}
\begin{align}
\tau_1\left(z\right)=&
-\frac{\eta  x^2 \left(\eta ^2-4 \gamma ^2\right) \sin (2 \eta  z)}{2\left[4\gamma^2+\eta^2+\left(\eta^2-4\gamma^2\right) \cos(2\eta  z)\right] }
+\frac{\varepsilon ^3 \eta  x \tan (\eta  z) \sec (\eta  z) \left[\eta ^2-4 \gamma ^2 \tan ^2(\eta  z)\right]}{2 \left[\eta ^2+4 \gamma ^2 \tan ^2(\eta  z)\right]^2}
\nonumber \\ &
-\frac{\varepsilon ^6 \eta  \sin ^3(\eta  z) \cos (\eta  z) \left[-12 \gamma ^2+\eta ^2+\left(12 \gamma ^2+\eta ^2\right) \cos (2 \eta  z)\right]}{3 \left[4 \gamma ^2+\eta ^2+\left(\eta ^2-4 \gamma ^2\right) \cos (2 \eta  z)\right]^3}
\nonumber \\ &
+\frac{\pi }{4}-\frac{1}{2} \arg \left[-\frac{2 \gamma  \tan (\eta  z)}{\eta }+i\right],\\
\tau_2\left(z\right) =&-\frac{2 \gamma  \eta^2}{4 \gamma^2+\eta^2+\left(\eta^2-4 \gamma^2\right) \cos(2\eta z)},\\
\tau_3\left(z\right)=&\frac{2 \gamma  \varepsilon ^3 \eta ^2  \tan ^2(\eta  z) \sec (\eta  z)}{\left[\eta ^2+4 \gamma ^2 \tan ^2(\eta  z)\right]^2},\\
\tau_4\left(z\right)=&-\frac{8 \gamma  \varepsilon ^6 \sin ^4(\eta  z) \left[-4 \gamma ^2+3 \eta ^2+\left(4 \gamma^2+3 \eta ^2\right) \cos (2 \eta  z)\right]}{12 \left[4 \gamma ^2+\eta ^2+\left(\eta ^2-4 \gamma ^2\right) \cos (2 \eta  z)\right]^3}-\frac{1}{4} \ln \left[\frac{4 \gamma ^2 \tan ^2(\eta  z)}{\eta ^2}+1\right]
\nonumber \\ &
-\frac{1}{2} \ln \left[ \cos(\eta z)\right] ,\\
\rho_1\left(z\right)=&\frac{\varepsilon  \eta^2 \sec (\eta  z)}{\eta ^2+4 \gamma ^2 \tan ^2(\eta  z)}
-i \frac{2 \gamma  \varepsilon  \eta  \tan (\eta  z) \sec (\eta  z)}{\eta ^2+4 \gamma ^2 \tan ^2(\eta  z)},\\
\rho_2\left(z\right)=&-\frac{\varepsilon ^4 \sin ^2(\eta  z) \left[-4 \gamma ^2+\eta ^2+\left(4 \gamma ^2+\eta ^2\right) \cos (2 \eta  z)\right]}{2 \left[4 \gamma ^2+\eta ^2+\left(\eta ^2-4 \gamma ^2\right) \cos (2 \eta  z)\right]^2}
\nonumber \\ &
+i\frac{4 \gamma  \varepsilon ^4 \eta  \sin ^3(\eta  z) \cos (\eta  z)}{\left[4 \gamma ^2+\eta ^2+\left(\eta ^2-4 \gamma ^2\right) \cos (2 \eta  z)\right]^2}.
\end{align}
\end{subequations}
Again, we have split the exponential in order to show that the first factor is just a phase. \\
These propagated fields can be substituted in the original paraxial equation, Eq. \eqref{ecschcua}, to prove directly that effectively they are the solutions. Furthermore, it is obvious that they satisfy the initial conditions. In Appendix \ref{feynmanprop}, we also present how to find the solution \eqref{psicuaexpai} using the Feynman propagator.\\
In Fig. \ref{3DIntCua}, we show the intensity of the fields expressed in \eqref{psicuaexpai} and \eqref{psicuagauai} as functions of $x$ and $z$. 
We already mentioned that the propagator \eqref{propecschcua} is in fact the fractional Fourier transform. So, for some values of $z$, it becomes the inverse Fourier transform and for others it is just the identity. It is important to note that even in places where the functions that define the propagator have singularities, the field is perfectly defined and continuous.\\

\section{The Bohm potential}
In 1979, Berry and Balazs \cite{Berry1979} found that a free quantum wavepacket exhibits unexpected accelerating features. This surprising theoretical result has now been experimentally tested both in optical and quantum mechanical settings using light and electrons beams \cite{ChavezCerda2011,Impens2020,Siviloglou2007,Christodoulides2004,Patsyk2018,Voloch_Bloch_2013}, respectively. In Ref.~\cite{Hojman12020}, it has been proved that this, as well as other unusual results, occurs in (classical or quantum) wave equations because their dispersion relations differ from the associated point-like dynamics dispersion relations due to the presence of the Bohm potential $V_\mathrm{B}(x,z)$, as defined by 
\begin{equation}\label{bohmpotential}
V_\mathrm{B}(x,z)=-\frac{1}{2 A(x,z)}\frac{\partial^2 A(x,z)}{\partial x^2},
\end{equation}
where $ A(x,z)$ is defined in terms of $\psi(x,z)$ as
\begin{equation}\label{0210}
A(x,z)=\left[ \psi(x,z) \psi^*(x,z)\right]^{1/2}.
\end{equation}
In this way, a generic complex solution $\psi(x,z)$ to the one-dimensional paraxial equation \eqref{ecschcua} is  written in terms of two real functions, an amplitude $A(x,z)$ and a phase $S(x,z)$,
\begin{equation}
  \psi(x,z)=A(x,z) e^{i S(x,z)}.
\end{equation}
The Bohm potential plays an essential role in quantum mechanics and in optics because it changes the classical dynamics (either the Hamilton-Jacobi or eikonal equations) by modifying the potential $V(x,t)$ in mechanics or the optical potential $V_{\mathrm{opt}}(x,z)$ in optics; the optical potential is defined by \cite{Marte1997}
\begin{equation}
V_{\mathrm{opt}}(x,z)=-\frac{n^2(x) -n_0^2}{2 n_0^2},
\end{equation}
where $n_0$ is the constant bulk refractive index and $n(x)$ is the GRIN index.\\
In other words, the dispersion relation for the eikonal equation differs from the paraxial equations dispersion relation due to the addition of the Bohm potential,
\begin{equation}
V_\mathrm{TotalOpt}(x,z)=V_\mathrm{opt}(x,z)+ V_\mathrm{B}(x,z).
\end{equation}
The eikonal equation associated to \eqref{ecschcua} gets an extra term (as compared to the ray optics limit) $V_\mathrm{B}(x,z)$, the Bohm potential, which appears as a modification to the GRIN medium refractive index in a Pythagorean or pseudo-Pythagorean  fashion, depending on the relative signs of (the squares of) the refractive index and the Bohm refractive index,
\begin{equation}\label{ntotal}
\left[ n_\mathrm{Total}(x,z)\right]^2=n^2(x)+ n_\mathrm{B}^2(x,z),
\end{equation}
in the ray optics limit. The Bohm modification to the GRIN index, $n_\mathrm{B}(x,z)$, is given by
\begin{equation}\label{nbohml}
\left[ n_\mathrm{B}(x,z)\right] ^2=-2V_\mathrm{B}(x,z).
\end{equation}
The Bohm potential may be computed directly considering \eqref{0210} to get the amplitude or by using the (logarithmic) derivative of the intensity $\rho(x,z)=\psi(x,z) \psi^*(x,z)$ as
\begin{equation}\label{vbohmrho}
V_\mathrm{B}(x,z)=-\frac{1}{4}\left[\frac{\rho'(x,z)}{\rho(x,z)}\right]'-\frac{1}{8}\left[\frac{\rho'(x,z)}{\rho(x,z)}\right]^2,
\end{equation}
where we have used the prime to denote derivative with respect to $x$.

\subsection{Bohm potential for the linear potential}
\begin{figure}[H]
\centering
\subfloat[Initial condition $\psi(x,z=0)=\exp(\gamma x)\textrm{Ai}(\epsilon x)$ with parameters $\gamma=0.25, \; \epsilon=2.0$ and $\eta=1.0$]
{\includegraphics[width=0.45\textwidth]{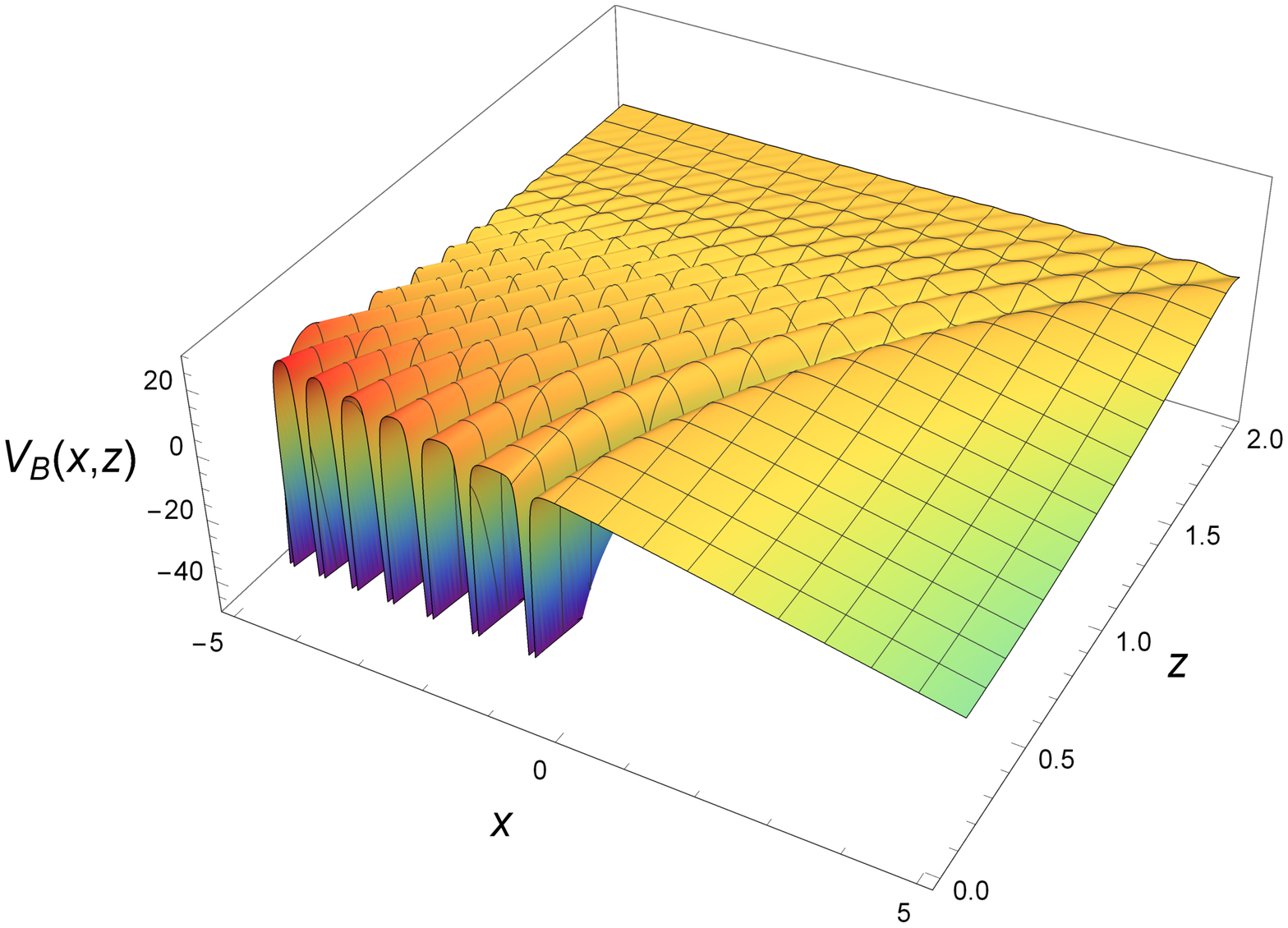}}
\hspace{14 pt}
\subfloat[Initial condition $\psi(x,z=0)=\exp(-\gamma x^2)\textrm{Ai}(\epsilon x)$ with parameters $\gamma=0.1, \; \epsilon=2.0$ and $\eta=2.0$]
{\includegraphics[width=0.45\textwidth]{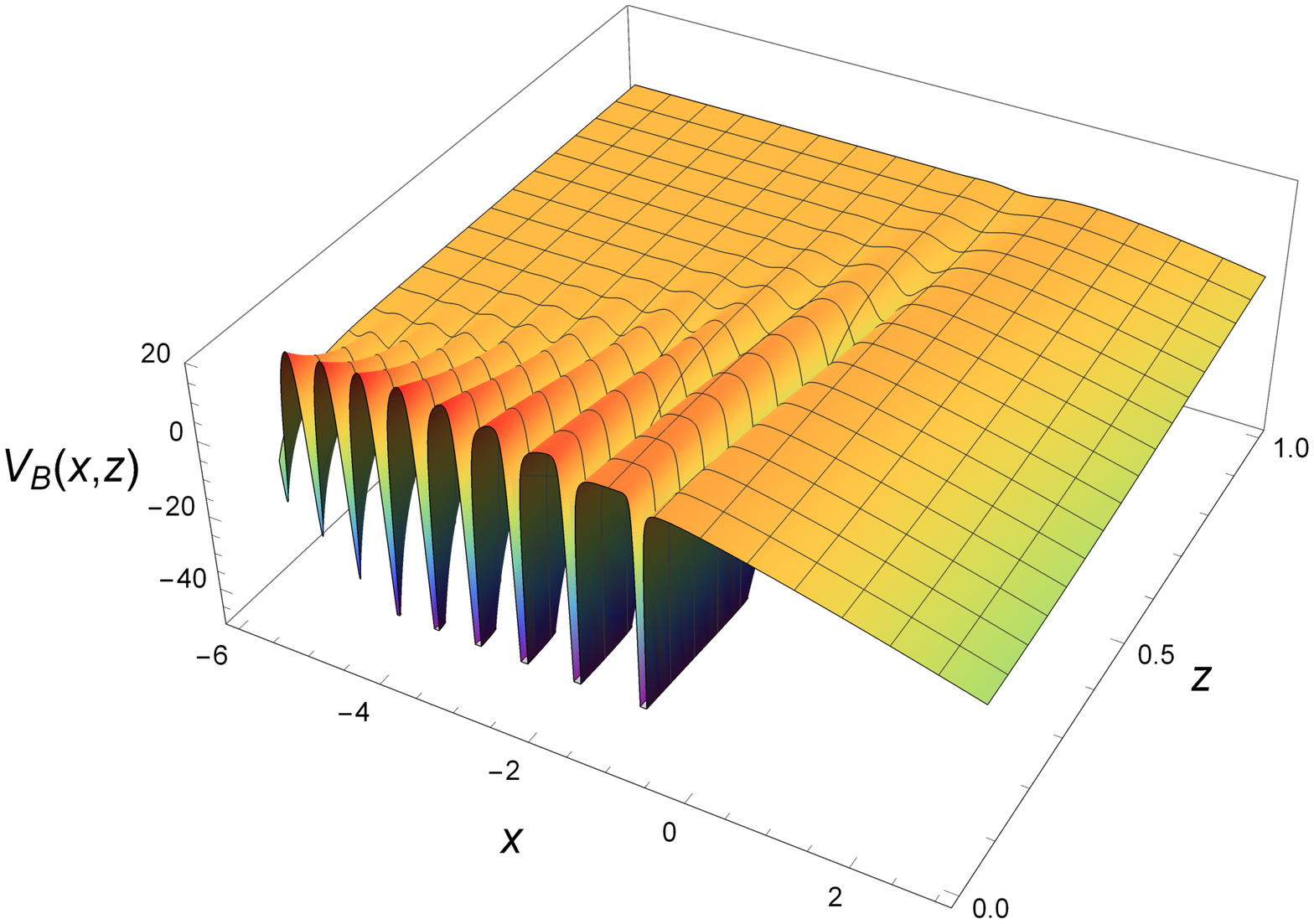}}
\caption{The Bohm potential of the linear potential for the two different initial conditions \label{3DVBLin}}
\end{figure}

Using Eq. \eqref{vbohmrho} for the Bohm potential and the solution of the paraxial equation for a linear GRIN medium with initial condition $\psi(x,z=0)=e^{\gamma x} \mathrm{Ai}(\epsilon x)$, Eq. \eqref{psilinexpai}, we find
\begin{align}\label{vblinexpai}
V_\mathrm{B}(x,z)=&-\frac{\gamma^2}{2}-\frac{\varepsilon^3 x}{2}+\frac{\varepsilon^3 z^2}{4}  \left(\frac{\varepsilon ^3}{2}-\eta \right)
-\frac{\gamma  \varepsilon}{2} \left[\frac{\text{Ai}'\left(y(x)\right)}{\text{Ai}\left(y(x)\right)}+\frac{\text{Ai}'(y^*(x))}{\text{Ai}(y^*(x))}\right]
\nonumber \\ &
+\frac{\varepsilon ^2}{8} \left[\frac{\text{Ai}'(y(x))}{\text{Ai}(y(x))}-\frac{  \text{Ai}'\left(y^*(x)\right)}{\text{Ai}\left(y^*(x)\right)}\right]^2,
\end{align}
where
\begin{equation}
y(x)=\varepsilon  x-\frac{1}{4} \varepsilon ^4 z^2+\frac{1}{2} \varepsilon  \eta  z^2+i \gamma  \varepsilon  z.
\end{equation}
Note that the Bohm potential has discontinuities for $x<-1$ and it becomes continuous for $x>-1$, due to the behavior of the Airy functions. For $x<-1$, the Airy functions behaves as sine and cosine functions so its derivative behaves as a tangent function. We also note that the Bohm potential becomes linear with the propagation.\\   
In the case of the Gaussian initial condition $\psi(x,z=0)=e^{-\gamma x^2} \mathrm{Ai}(\epsilon x)$, Eq. \eqref{psilingauai}, we find
\begin{align}\label{vblingaussai}
& V_\mathrm{B}(x,z)=
\frac{\varepsilon ^3 x \left(12 \gamma ^2 z^2-1\right)}{2 \left(4 \gamma ^2 z^2+1\right)^3}
-\frac{\varepsilon^2}{4 (i-2\gamma z)^2}
\frac{\text{Ai}'^2(y(x))}{\text{Ai}^2(y(x))}
-\frac{\varepsilon^2}{4 (i+2\gamma z)^2}
\frac{\mathrm{Ai}'^2\left(y^*(x)\right)}{\text{Ai}^2\left(y^*(x)\right)}
\nonumber \\ &
-\frac{1}{8} \left[-\frac{4 \gamma  x}{4 \gamma ^2 z^2+1}
-\frac{2 \gamma  z^2 \left(-2 \varepsilon ^3+\eta +4 \gamma ^2 \eta  z^2\right)}{\left(4 \gamma ^2 z^2+1\right)^2}
+i\frac{\varepsilon}{i-2\gamma z}
\frac{\text{Ai}'(y(x))}{\text{Ai}(y(x))}
+i\frac{\varepsilon}{i+2\gamma z}
\frac{\text{Ai}'\left(y^*(x)\right)}{\text{Ai}\left(y^*(x)\right)}
\right]^2
\nonumber \\ &
+\frac{8 \gamma +512 \gamma ^7 z^6+16 \gamma ^4 \varepsilon ^3 z^6 \left(\varepsilon ^3+6 \eta \right)+384 \gamma ^5 z^4-8 \gamma ^2 \varepsilon ^3 z^4 \left(3 \varepsilon ^3-2 \eta \right)+96 \gamma ^3 z^2+\varepsilon ^3 z^2 \left(\varepsilon ^3-2 \eta \right)}{8 \left(4 \gamma ^2 z^2+1\right)^4},
\end{align}
where
\begin{equation}
y(x)=\frac{-\varepsilon x (4+8 i \gamma z)+\varepsilon z^2 \left(\varepsilon ^3-2 \eta -4 i \gamma  \eta  z\right)}{4 (i-2\gamma z)^2}.
\end{equation}
In Fig. \ref{3DVBLin}, we plot the Bohm potential as function of $x$, the transverse coordinate, and of $z$, the propagation distance for the two initial conditions that we have considered.
Both graphics start at $z=0.1$ in order to avoid the discontinuities mentioned above that the Bohm potential has at $z=0$.

\subsection{Bohm potential for the quadratic potential}
\begin{figure}[H]
\centering
\subfloat[Initial condition $\psi(x,z=0)=\exp(\gamma x)\textrm{Ai}(\epsilon x)$ with parameters $\gamma=1.0, \; \epsilon=2.0$ and $\eta=1.0$]
{\includegraphics[width=0.45\textwidth]{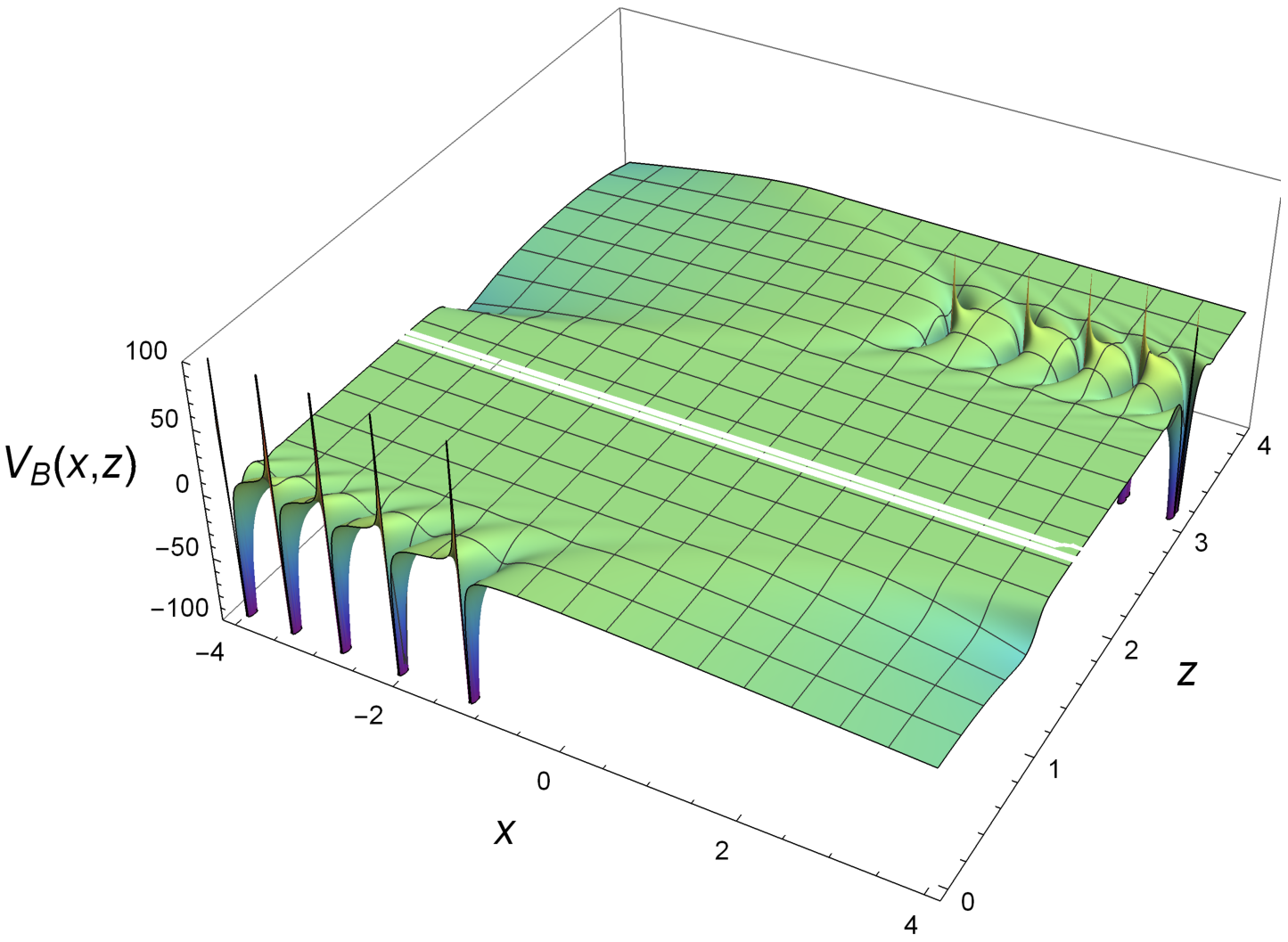}}
\hspace{14 pt}
\subfloat[Initial condition $\psi(x,z=0)=\exp(-\gamma x^2)\textrm{Ai}(\epsilon x)$ with parameters $\gamma=1.0, \; \epsilon=2.0$ and $\eta=1.0$]
{\includegraphics[width=0.45\textwidth]{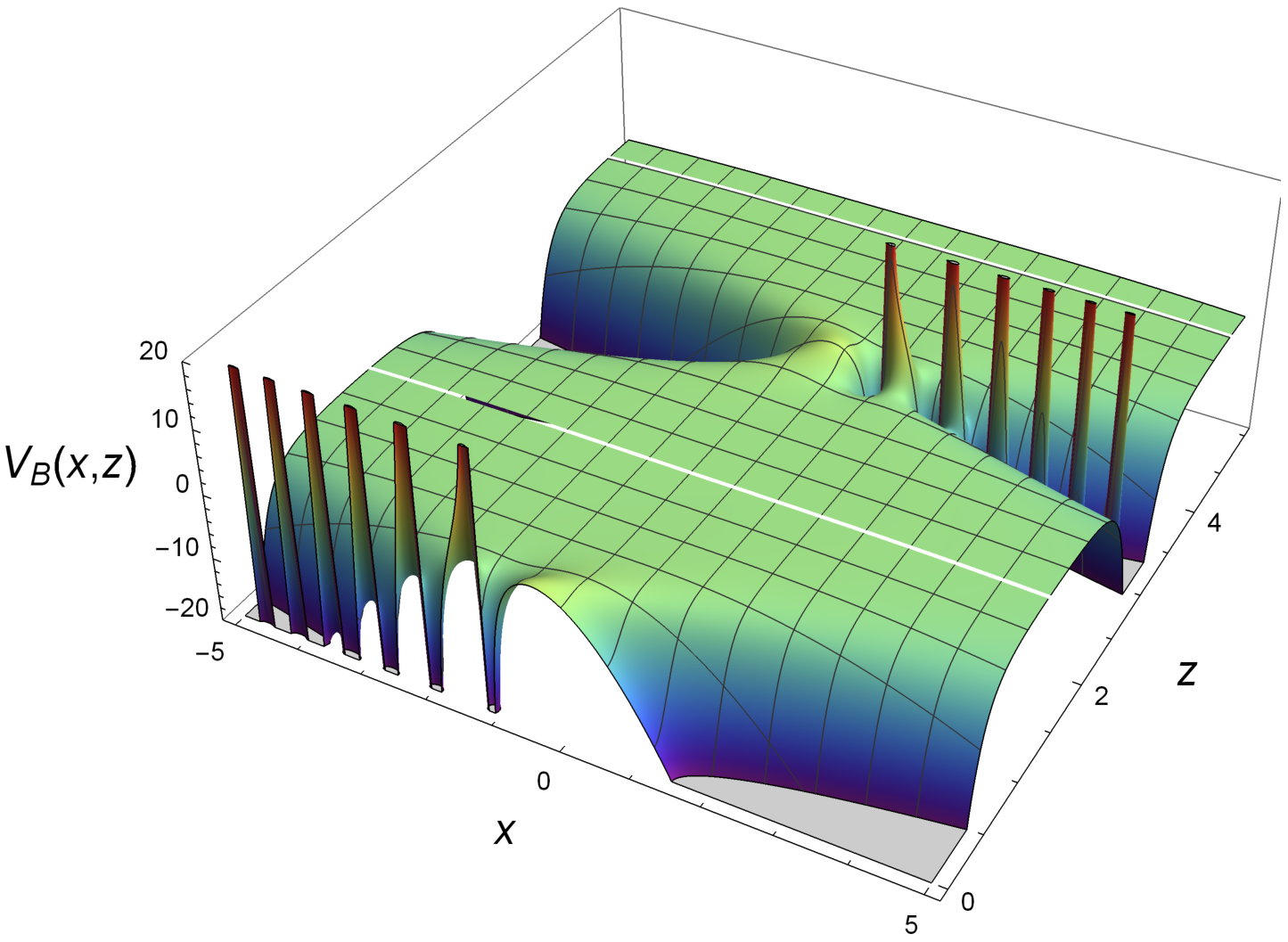}}
\caption{The Bohm potential of the quadratic potential for the two different initial conditions \label{3DVBCua}}
\end{figure}

In the case of the quadratic potential with an initial beam $\psi(x,z=0)=\exp(\gamma x) \textrm{Ai}(\epsilon x)$, the Bohm potential is given by
\begin{align}\label{vbcuaexpai}
& V_\mathrm{B}(x,z)=
\frac{\varepsilon^2\sec^2(\eta  z)}{4}
\left[ \frac{ \text{Ai}'^2\left(y^*(x,z)\right)}{ \text{Ai}^2\left(y^*(x,z)\right)}+\frac{ \text{Ai}'^2(y(x,z))}{ \text{Ai}^2(y(x,z))}\right] 
\nonumber \\ &
-\frac{\sec^2(\eta z)}{8} \left[\varepsilon \frac{\text{Ai}'\left(y^*(x,z)\right)}{\text{Ai}\left(y^*(x,z)\right)}+\varepsilon \frac{\text{Ai}'(y(x,z)) }{\text{Ai}(y(x,z))}+2 \gamma \right]^2
+\frac{\varepsilon ^6 \tan ^2(\eta  z) \sec ^2(\eta  z)}{8 \eta ^2}
-\frac{1}{2} \varepsilon^3 \sec^3(\eta  z) x
,
\end{align}
where
\begin{equation}
y(x,z)=\varepsilon \sec (\eta  z) x
-\frac{\varepsilon^4 \tan ^2(\eta  z)}{4 \eta ^2}
+ i \frac{\gamma \varepsilon \tan(\eta z)}{\eta}.
\end{equation}
The Bohm potential for this case, Eq. \eqref{vbcuaexpai}, has singularities in $\eta z=n \frac{\pi}{2}$ with $n$ integer, due to the singularities of $\sec(\eta z)$.\\
In the case of the initial condition $\psi(x,z=0)=e^{-\gamma x^2} \mathrm{Ai}(\epsilon x)$ with propagated field Eq. \eqref{psicuagauai}, we find
\begin{align}\label{vbcuagaussai}
V_\mathrm{B}(x,z)=&-\frac{1}{8} \left[
\mu_1(x,z)+\mu_2(z)
\frac{\text{Ai}'(y(x,z))}{\text{Ai}(y(x,z))}
+\mu_2^*(z)
\frac{\text{Ai}'\left(y^*(x,z)\right)}{\text{Ai}\left(y^*(x,z)\right)}
\right]^2
\nonumber \\ &
+\frac{1}{4} \left[
\mu_3(x,z)+\mu_2^2(z)
\frac{\text{Ai}'^2(y(x,z))}{\text{Ai}^2(y(x,z))}
+\mu_2^{* 2}(z)
\frac{\text{Ai}'^2\left(y^*(x,z)\right)}{\text{Ai}^2\left(y^*(x,z)\right)}
\right]
\end{align}
where
\begin{equation}
\mu_1(x,z)=\frac{4 \gamma  \varepsilon ^3 \eta ^2 \tan ^2(\eta  z) \sec (\eta  z)}{\left(\eta ^2+4 \gamma ^2 \tan ^2(\eta  z)\right)^2}-\frac{8 \gamma  \eta ^2 x}{4 \gamma ^2+\eta ^2+\left(\eta ^2-4 \gamma ^2\right) \cos (2 \eta  z)}, 
\end{equation}
\begin{equation}
\mu_2(z)=\frac{\varepsilon  \eta  \sec (\eta  z)}{\eta +2 i \gamma  \tan (\eta  z)},
\end{equation}
\begin{align}
&\mu_3(x,z)=
-\frac{\varepsilon ^3 \eta ^4 x \sec ^3(\eta  z) \left[\eta ^2-12 \gamma ^2 \tan ^2(\eta  z)\right]}{2 \left[\eta ^2+4 \gamma ^2 \tan ^2(\eta  z)\right]^3}
\nonumber \\ &
+\frac{5120 \gamma ^7 \eta ^2+768 \gamma ^5 \eta ^4+160 \gamma ^4 \varepsilon ^6 \eta ^2+192 \gamma ^3 \eta ^6-48 \gamma ^2 \varepsilon ^6 \eta ^4+80 \gamma  \eta ^8+2 \varepsilon ^6 \eta ^6}{16 \left[4 \gamma ^2+\eta ^2+\left(\eta ^2-4 \gamma ^2\right) \cos (2 \eta  z)\right]^4}
\nonumber \\ &
+\frac{\left(-7680 \gamma ^7 \eta ^2-384 \gamma ^5 \eta ^4-240 \gamma ^4 \varepsilon ^6 \eta ^2+96 \gamma ^3 \eta ^6+24 \gamma ^2 \varepsilon ^6 \eta ^4+120 \gamma  \eta ^8+\varepsilon ^6 \eta ^6\right) \cos (2 \eta  z)}{16 \left[4 \gamma ^2+\eta ^2+\left(\eta ^2-4 \gamma ^2\right) \cos (2 \eta  z)\right]^4}
\nonumber \\ &
+\frac{\left(3072 \gamma ^7 \eta ^2-768 \gamma ^5 \eta ^4+96 \gamma ^4 \varepsilon ^6 \eta ^2-192 \gamma ^3 \eta ^6+48 \gamma ^2 \varepsilon ^6 \eta ^4+48 \gamma  \eta ^8-2 \varepsilon ^6 \eta ^6\right) \cos (4 \eta  z)}{16 \left[4 \gamma ^2+\eta ^2+\left(\eta ^2-4 \gamma ^2\right) \cos (2 \eta  z)\right]^4}
\nonumber \\ &
+\frac{\left(-512 \gamma ^7 \eta ^2+384 \gamma ^5 \eta ^4-16 \gamma ^4 \varepsilon ^6 \eta ^2-96 \gamma ^3 \eta ^6-24 \gamma ^2 \varepsilon ^6 \eta ^4+8 \gamma  \eta ^8-\varepsilon ^6 \eta ^6\right) \cos (6 \eta  z)}{16 \left[4 \gamma ^2+\eta ^2+\left(\eta ^2-4 \gamma ^2\right) \cos (2 \eta  z)\right]^4}
\end{align}
and
\begin{equation}
y(x,z)=\xi_1\left(z\right) x+\xi_2\left(z\right),
\end{equation}
with
\begin{subequations}
\begin{align}
\xi_1\left(z\right)=&\frac{\varepsilon  \eta ^2 \sec (\eta  z)}{\eta ^2+4 \gamma ^2 \tan ^2(\eta  z)}
-i\frac{2 \gamma  \varepsilon  \eta \tan (\eta  z) \sec (\eta  z)}{\eta ^2+4 \gamma ^2 \tan ^2(\eta  z)}
  ,\\
\xi_2\left(z\right)=&
-\frac{\varepsilon ^4 \sin ^2(\eta  z) \left(-4 \gamma ^2+\eta ^2+\left(4 \gamma ^2+\eta ^2\right) \cos (2 \eta  z)\right)}{2 \left[4 \gamma ^2+\eta ^2+\left(\eta ^2-4 \gamma ^2\right) \cos (2 \eta  z)\right]^2}
+i \frac{4 \gamma  \varepsilon ^4 \eta  \sin ^3(\eta  z) \cos (\eta  z)}{\left[4 \gamma ^2+\eta ^2+\left(\eta ^2-4 \gamma ^2\right) \cos (2 \eta  z)\right]^2} .
\end{align}
\end{subequations}
In Fig. \ref{3DVBCua}, we plot the Bohm potential as function of $x$, the transverse coordinate, and of $z$, the propagation distance for the two considered initial conditions. 
The singularities at $\eta z=n \frac{\pi}{2}$, with $n$ integer, are excluded from the graphics.\\

\section{Conclusions}
We have shown that an Airy beam keeps its form during propagation in a GRIN medium, and, when made finite by its multiplication with an exponential function, it still maintains the form, however, with a complex argument. Moreover, given the structure of equations (\ref{propecschlin2}) and (\ref{propecschcua}), that have an exponential of free propagation, it may be inferred that any beam that keeps its form during free propagation, also keeps it during propagation in indexes of refraction that produce linear and quadratic potentials.\\
It is interesting to remark that the propagation of this kind of beams gives rise to (non-vanishing) Bohm potentials which essentially means that the total refractive index ${n}_\textrm{Total}(x,z)$, Eq. \eqref{ntotal}, for ray optics is non-trivially different from the material GRIN index and, in fact, may depend on both $x$ and $z$, even if the GRIN index is $z$ independent, as it happens in the cases presented here.

\section{Acknowledgments}
The authors want to thank Dr. Víctor Arrizón from the National Institute of Astrophysics, Optics and Electronics (INAOE) for his participation in the first discussions of this work and for his valuable comments.

\appendix

\section{Factorization of the propagator}\label{apa}
Let us consider the paraxial equation
\begin{equation}\label{a10010}
i\frac{\partial \psi}{\partial z}=-\frac{1}{2}\frac{\partial^2 \psi}{\partial x^2}+\frac{\eta^2x^2}{2}\psi,
\end{equation}
where $\psi$ is a function of $x$ and $z$, and $\eta$ is a parameter of the GRIN medium.\\
We define the operators
\begin{equation}\label{a10020}
\hat{q}=\eta x, \qquad   \hat{p}=-i\frac{\partial}{\partial x},
\end{equation}
in terms of which the paraxial equation becomes
\begin{equation}\label{a10030}
i\frac{\partial \psi}{\partial z}=-\frac{1}{2}\left( \hat{p}^2
+\hat{q}^2\right) \psi,
\end{equation}
and its propagator is
\begin{equation}\label{a10040}
\hat{U}(z)=\exp\left[-\frac{i}{2}\left(\hat{p}^2+\hat{q}^2\right) z \right].
\end{equation}
In this Appendix, we show that it is possible to find functions $f\left(z\right) $, $g\left(z\right)$ and $h\left(z\right)$ such that
\begin{equation}\label{a10050}
\hat{U}(z)=\exp\left[-\frac{i}{2} f\left(z\right)\hat{q}^2\right] 
\exp\left[-\frac{i}{2} g\left(z\right)\left( \hat{q}\hat{p}+\hat{p}\hat{q}\right) \right] 
\exp\left[-\frac{i}{2} h\left(z\right)\hat{p}^2\right];
\end{equation}
as $\hat{U}(0)=\hat{I}$, it is clear that those functions must satisfy the initial conditions $f\left(0\right)=0, \,
g\left(0\right)=0, \, h\left(0\right)=0$.\\
We start deriving the propagator \eqref{a10040} and the proposal \eqref{a10050} with respect to $z$, and we make all the required transformations to find the coupled ordinary differential equations system
\begin{subequations}
\begin{align}
\exp\left(2\eta g\right) \frac{dh}{dz}&=1, \label{a1016a0} \\
\eta^2 f^2 \exp\left(2\eta g\right)  \frac{dh}{dz}+2\eta f  \frac{dg}{dz}+\frac{df}{dz}&=1, \label{a1016b0}\\
\eta  f \exp\left(2\eta g\right) \frac{dh}{dz}+\frac{dg}{dz}&=0, \label{a1016c0}
\end{align}	
\end{subequations}
with the corresponding initial conditions. This system is easy to solve and we find
\begin{subequations}
\begin{align}
f\left(z\right)=&\frac{1}{\eta}\tan\left(\eta z\right),
\\
g\left( z\right) =&\frac{1}{\eta}\ln\left[\cos\left(\eta z \right) \right],
\\
h\left( z\right) =&f\left( z\right).
\end{align}
\end{subequations}

\section{A different way to get the solution}\label{feynmanprop}
The wavefunction $\psi(x_f, t_f)$ for any point $x_f$ and any time $t_f$ which solves the Schrödinger equation for any potential may be written in terms of an integral which involves the Feynman propagator $K(x_f, x_i, t_f,t_i)$ and an arbitrary initial value wavefunction $\psi(x_i, t_i)$ for an initial point $x_i$ and time $t_i$, 
\begin{equation}\label{prop1}
\psi(x_f, t_f)\ =\ \int_{- \infty}^{\infty} dx_i  \ K(x_f, x_i, t_f,t_i)\ \psi(x_i, t_i)\ , 
\end{equation}
\noindent provided  $K(x_f, x_i, t_f,t_i)$ is known for a given potential.\\
The Feynman propagator for the one-dimensional harmonic oscillator is
\begin{align}\label{propho}
K(x_f, x_i, t_f,t_i)=&\left\lbrace \ \frac{m \omega}{2 \pi i \hslash\ \sin\left[ \omega (t_f-t_i)\right] }\ \right\rbrace ^{1/2}
\nonumber \\ &
\exp\left(\ -\ \frac{m \omega\left\lbrace ({x_f}^2+{x_i}^2)\ \cos\left[ \omega(t_f-t_i)\right] -2 x_f x_i\right\rbrace }{2 i    \hslash  \ \sin\left[ \omega (t_f-t_i)\right] }\right)\ .
\end{align}
If we choose as initial wavefunction $\psi(x_i, t_i)$
\begin{equation}\label{psii}
\psi(x_i, 0)\ = \exp(\gamma x_i)\ Ai(\epsilon x_i)
\end{equation}
and we make all the integration's, we get
\begin{align}
\psi(x,z)=&
\frac{1}{2 \pi}\exp\left( i \gamma^2 f-i \frac{2\epsilon^6 f^3}{3}-\frac{g}{2}-2\epsilon^3 \gamma f^2\right)
\exp\left[ -i\eta^2 f x^2+\left(\gamma+i\epsilon^3 f\right)\exp\left(-g\right)x\right]  
\nonumber \\ &
\mathrm{Ai}\left(\epsilon \exp\left(-g\right)x+2i\epsilon\gamma f-\epsilon^4 f^2\right).
\end{align}
which coincides with Eq, \eqref{psicuaexpai} apart from an irrelevant $\frac{1}{2 \pi}$ factor. Similar calculations can be performed for the other cases.

\end{document}